\documentclass[fleqn,usenatbib]{mnras}
\usepackage{newtxtext,newtxmath}
\usepackage[T1]{fontenc}
\usepackage{ae,aecompl}
\usepackage{graphicx}
\usepackage{amsmath}
\usepackage{amssymb}

\hypersetup{
    colorlinks=true,     
    urlcolor=blue
}

\title[Segmentation of coronal holes]{Segmentation of coronal holes in solar disk images with a convolutional neural network}

\author[E.A. Illarionov et al.]{
Egor A. Illarionov,$^{1, 3}$\thanks{E-mail: egor.mypost@gmail.com}
Andrey G. Tlatov,$^{2, 3}$
\\
$^{1}$Moscow State University, Moscow, 119991, Russia\\
$^{2}$ Kislovodsk Mountain Astronomical Station of the Pulkovo Observatory, Kislovodsk, 357700, Russia \\
$^{3}$ Kalmyk State University, Elista, 358000, Russia
}

\date{Accepted XXX. Received YYY; in original form ZZZ}

\pubyear{2018}

\begin{document}
\label{firstpage}
\pagerange{\pageref{firstpage}--\pageref{lastpage}}
\maketitle

\begin{abstract}
Current coronal holes segmentation methods typically rely on image thresholding and require non-trivial image pre- and post-processing.
We have trained a neural network that accurately isolates CHs from SDO/AIA~193~{\AA}
solar disk images without additional complicated steps. 
We compare results with publicly available catalogues of CHs and demonstrate stability of the 
neural network approach. In our opinion, this approach can outperform hand-engineered 
solar image analysis and will have a wide application to solar data. In particular, we 
investigate long-term variations of CH indices within the solar cycle 24 and observe increasing of CH areas in about three times from minimal values in the maximum of the solar cycle to maximal 
values during the declining phase of the solar cycle.
\end{abstract}


%%%%%%%%%%%%%%%%%%%%%%%%%%%%%%%%%%%%%%%%%%%%%%%%%%

%%%%%%%%%%%%%%%%% BODY OF PAPER %%%%%%%%%%%%%%%%%%

\section{Introduction}

Coronal holes (CHs) are regions of open magnetic fields in the solar corona.
They are typically associated with high speed solar wind streams \citep[e.g.][]{Altschuler1972, Krieger1973}
and provide an estimation of solar wind parameters and corresponding geomagnetic effects 
\citep[e.g.][]{Nolte1976, Robbins2006, Obridko2009, Abramenko2009, Rotter2015}. 
An accurate CHs segmentation procedure is essential for more detailed investigation 
of this relationship and implementation of space weather forecasting routines.

CHs are best seen by eye in images in the 193~{\AA} wavelength. They appear as dark regions 
due to their lower density and temperature in contrast to surrounding atmosphere \citep[see][]{priest2014magnetohydrodynamics}.
CHs have irregular profile and may occupy more than 10\% of visible solar hemisphere.

For segmentation of CHs one typically applies rule-based approaches that include image thresholding
and region growth steps. For example, \citet{Krista2009} identify CHs using a histogram-based
intensity thresholding. The Automatic Solar Synoptic Analyzer \citep[ASSA;][]{ASSA} also
includes thresholding at some fraction of median pixel value.
Morphological image analysis, thresholding and smoothing was
applied by \citet{2005ASPC..346..261H} for CH detection in solar spectroheliograms.
\citet{Scholl2008} apply contrast enhancement in a multi-passband detection method
in 171~{\AA}, 195~{\AA}, and 304~{\AA} passbands. Similarly, multi-thermal intensity cut is a key step
in the Coronal Hole Identification via Multi-thermal Emission Recognition Algorithm \citep[CHIMERA;][]{2018JSWSC...8A...2G}.
The Spatial Possibilistic Clustering Algorithm (SPOCA) requires a proper intensity normalization to
apply fuzzy clustering method for image segmentation \citep{SpoCA2009, SpoCA2014}. 

While contrast enhancement, thresholding and geometrical 
considerations were massively exploited in computer vision for may years in a number of domains, more and more 
recent publications demonstrate that computer vision algorithms based on neural networks
can provide significantly better results in object detection
in comparison to conventional hand-engineered algorithms (see e.g.\citet{ChenPKMY14} for a review).
In application to solar data, \citet{Baso2018} presented deep learning techniques for enhancing SDO/HMI images,
while \citet{Ramos2017} trained a neural network to estimate horizontal velocities at the solar surface.
We failed to find any publication relevant to neural networks usage for active regions segmentation, although the amount of available data for neural network training is more than enough.

In our opinion, investigation of neural networks with respect to segmentation of solar
active regions allows some possibilities beyond the capacity of hand-engineered algorithms.
The point is that neural networks can learn a methodology of active regions isolation
that was used e.g. at some observatory or by particular astronomer. Applying the trained network
to a larger time interval one can create an extended dataset of active regions isolated
by methodology similar to original one. This trick can hardly be done by manual
threshold levels adjustment since many aspects of individual image perception are difficult to formalize explicitly.
Such an extended dataset would be interesting in respect to e.g. calibration
of sunspot number series (see \citet{Clette2014} for review of the problem).

This paper is aimed to present an approach based on convolutional neural networks (CNNs)
for segmentation of coronal holes in solar disk images and attract interest in
application of this technique to wider scope of astrophysical problems.
We demonstrate that CNN can outperform hand-engineered methods of CH detection
and investigate properties of obtained CHs within the solar cycle 24.

\begin{figure*}
	\includegraphics[width=\textwidth]{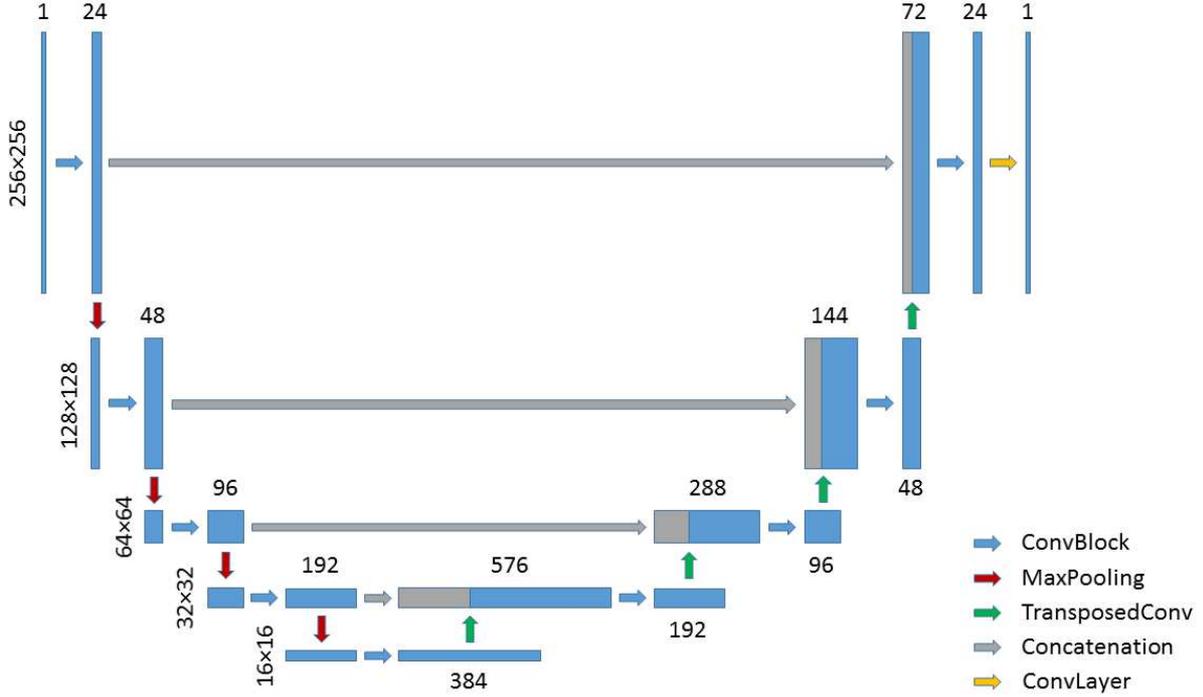}
    \caption{Neural network architecture for CHs segmentation. Each box corresponds to a multi-channel feature map. Width of boxes and numbers above or below boxes represent its channel dimension, while height of each box and numbers to the left of boxes represent its spatial dimensions. Color arrows correspond to different operations annotated in the figure.}
    \label{fig:unet}
\end{figure*}

\section{Convolutional neural networks for image segmentation}

Convolutional neural network consists of an input and output layers and a number
of hidden layers. Hidden layers can be represented by convolutional, resizing, normalization 
and some other type of layers. The detailed definition of layers and an intuition behind them
can be found e.g. in \citet{Bishop:1995}, \citet{Li2014} or \cite{Goodfellow2016}.
The way layers are connected defines a  neural network architecture.
Training of neural networks consists in optimization of trainable variables
with respect to loss function between target values and output predictions of the model.

\subsection{Network architecture}
For segmentation of CHs we implemented an architecture similar to the U-Net architecture,
elaborated for biomedical image segmentation \citep[see][]{unet}.
It is worth noting that U-Net-like networks demonstrate superior results
in very different segmentation problems, in 2D or 3D. For example, in satellite
image analysis \citep{IglovikovMO17} or medical image analysis \citep{Iglovikov2017, Ching142760, CicekALBR16}.
These facts inspired us to apply U-Net-like network for segmentation of CHs.

Fig.~\ref{fig:unet} shows an architecture of the neural network that
we suggest for CHs segmentation. For simplicity we will
also call this network U-Net. 
The network input and output layers have the same
shape $256\times 256\times 1$. Input layer accepts grayscale image of solar disk resized 
to $256 \times 256$ pixels. Output layer produces a predicted segmentation map, i.e. an image of 
the same shape $256 \times 256$, where pixels assigned to CHs have values close to 1, 
while other pixels have values close to 0. Hidden layers are represented by operations named
ConvBlock, MaxPooling, TransposedConv, Concatenation and Conv and are explained below.

ConvBlock is a compact notation of two successive convolutional layers with activation
function called "exponential linear unit" \citep[ELU, see][]{ClevertUH15}. 
Both convolutional layers have the same number of filters, these numbers are shown in Fig.~\ref{fig:unet}.
Note that we double number of filters in ConvBlocks from top to bottom of the network starting from 24.
Kernel size for convolutions is $3 \times 3$ 
(any other convolutions in our network also have kernel size $3 \times 3$).
Spatial resolution remains unchanged after convolutions due to zero padding and unitary stride.
In the end of ConvBlock we apply a dropout layer \citep{srivastava14a} with rate 0.1.
The dropout layer deactivates a certain set of neurons chosen at random with given rate during
training phase. This helps the neural network to improve over-fit.   

MaxPooling layer reduces spatial resolution to a half using maximum function to summarize subregions.
Channel dimension remains unchanged. Size of the pooling window is 2 for both spatial axes, strides are also equal to 2.

TransposedConv is a convolutional layer with transposed gradient \citep[see][]{dumoulin2016guide} and 
ELU activation function. Due to strides equal to 2 for all spatial axes, the output spatial dimensions are doubled.
Sometimes this operation is called "deconvolution". Here we also apply dropout with rate 0.1.

Concatenation operation simply stacks two tensors along channel axis.

Conv layer is a convolution layer with sigmoid activation function $\sigma(x) = (1 + e^{-x})^{-1}$.
Sigmoid function ensures that the network output values are in [0, 1] range.
Recall that we interpret these values as probability of pixels to belong to CH.

Intuition behind the U-net architecture can be understood as follows. Left branch of
the network compress an input image to extract context information. Connections between
left and right branches pass context information to improve its localization during
decompression. 

\subsection{Dataset}
A dataset for CH segmentation consists of pairs
of solar disk images and corresponding segmentation maps.
We exploit one image per day in $1024\times 1014$ resolution from the Solar Dynamic Observatory (SDO) Atmospheric Imaging Assembly \citep[AIA;][]{Lemen2012} 193~{\AA} catalogue. It
covers time period from 2010 to the present time.
The catalogue is accessible e.g. at the Joint Science Operation
Center (JSOC; \href{http://jsoc.stanford.edu/}{http://jsoc.stanford.edu/}).
Segmentation maps are obtained from data archive of the
Kislovodsk Mountain Astronomical Station that is available and daily updated at \href{http://observethesun.com}{http://observethesun.com}.
Recall a CH processing algorithm for this data archive (we will refer to it as
the region growth algorithm).
 
CHs are obtained as a result of semi-automatic
procedure of processing SDO/AIA~193~{\AA} solar disk images. This procedure inherits the ones applied in \citet{Tlatov2014} for CHs identification from observations in the He~I~10830~{\AA} line made at Kitt Peak Observatory (from 1975 to 2003) and in the EUV~195~{\AA} wavelength with SOHO/EIT (from 1996 to 2012) with some improvements.

In the first step, the algorithm detects solar disk centre and radius.
Then, it computes pixels intensity distribution within the solar disk
and selects initial regions with pixel intensity below 0.4 quantile. Here it also detects the quite Sun level ($LQS$) as the mode of the distribution. Note that the threshold  parameter for initial regions, which is 0.4 by default, is controlled by the observatory data engineer and can be varied in order to achieve better agreement with visual expectation.
The next step consists of iterative region growth
from initial regions. At each iteration new pixels with intensity above $0.4 + 0.05 * i * LQS$ are merged into region, where $i$ is the current iteration step. Iteration stops when the region area growth rate rapidly leaps up. At this moment the region starts to occupy regions of the quite Sun.   

Obtained regions
with areas less than 2K millionth of a solar hemisphere (MSH) are filtered out. 
Finally, CH candidate
regions are inspected visually and the threshold parameter for initial regions is manually corrected if necessary.
The result containing coordinates of CHs boundaries and parameters of CHs (e.g. area and elongation) is added to the map of solar active
regions at \href{http://observethesun.com}{http://observethesun.com}.

Totally the dataset contains 2916 pairs of solar disk images and corresponding segmentation maps.
We divide it into train and test parts, where train part contains all pairs before 2017 (2385 items) and
test part contains all pairs starting from 2017 (513 items). Note that we do not mix dates in train and test subsets. The point is that otherwise the relative low day to day variability of CHs \citep[see e.g.][]{Zhang2003} will result in very similar items in the test and train subsets and thus will distort the model evaluation.

\subsection{Data preprocessing}
Image preprocessing 
consists of resizing of original $1024\times 1024$ images to shape $256\times 256$,
rescaling of pixel values to [0, 1] range from original 0, 1 ... 255 intensity values, 
rotation of each image at random angle in {0, 90, 180, 270} degrees and reversing along each spatial
axis with probability 0.5. Random transformations included in the preprocessing increase the amount
of training data (augment the dataset) and reduce over-fit \citep[see e.g.][]{Perez2017}.

Segmentation maps are given by images of shape $256\times 256$ resized from
original $1024\times 1024$ maps. Each pixel
is equal to 1 or 0 depending on whether it belongs to CH or not. 

\subsection{Training}
We use preprocessed solar disk images and corresponding segmentation maps to train the network
with the adaptive moment estimation optimization algorithm\citep[Adam;][]{KingmaB14} with learning rate 0.001.

Similarity measure
between predicted and target segmentation map (loss function for optimization algorithm)
is computed by a pixel-wise binary
cross-entropy defined as
\begin{equation}
loss(Y, \hat{Y}) = -\frac{1}{n^2}\sum\limits_{i, j = 0}^{n} y_{ij}\log\hat{y}_{ij} + (1 - y_{ij})\log(1 - \hat{y}_{ij})\,,
\end{equation}
where $Y = (y_{ij})$, $\hat{Y} = (\hat{y}_{ij})$ are target and predicted 
segmentation maps correspondingly.

Training procedure consists of iterations over the dataset with subsets of certain size
and updating trainable variable of the network after each iteration according
to given optimization algorithm. The subsets are called batches, one iteration over
the whole dataset is called epoch. Once the dataset is exhausted, we permute its
items at random and start a new epoch. Thus we obtain batches composed of random items and improve
convergence of the optimization algorithm \citep[see e.g.][]{Meng2017ConvergenceAO}.
As it was shown e.g. in \citet{batch2017}, larger batch sizes provide better classification or
segmentation results for neural networks. However, there is always a trade-off between batch size,
neural network size and and CPU/GPU memory limits. 
The U-Net network has about 6.2M of trainable variables.
We used the GeForce GTX 1060 graphics card with 6 Gb memory 
for experiments and batch size 20. Training time is about 15 minutes using . It is worth noting that the same experiments 
on CPU were more than 10 times as long for us.

\begin{figure}
\centering\includegraphics[width=\linewidth]{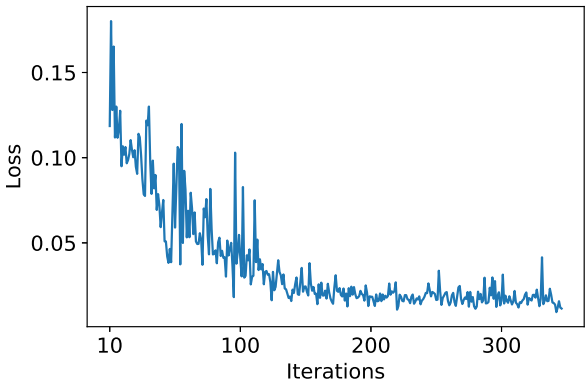}
    \caption{Loss function (binary cross-entropy) against iterations. First ten iterations are hidden due to large values of the loss function.}
    \label{fig:loss}
\end{figure}

Fig.~\ref{fig:loss} shows how the loss function varies with iterations.
We observe that 3 epochs (357 iterations) are enough to reach a plateau in the loss function, which means a convergence to some proper local minimum.
Averaged loss at the final epochs is equal to 0.018.

The source code necessary to reproduce the training procedure is provided in a public repository \href{https://github.com/observethesun/coronal_holes}{https://github.com/observethesun/coronal\_holes}. Python and TensorFlow are required to run the code.

\section{Results}
\begin{figure}
\includegraphics[width=\columnwidth]{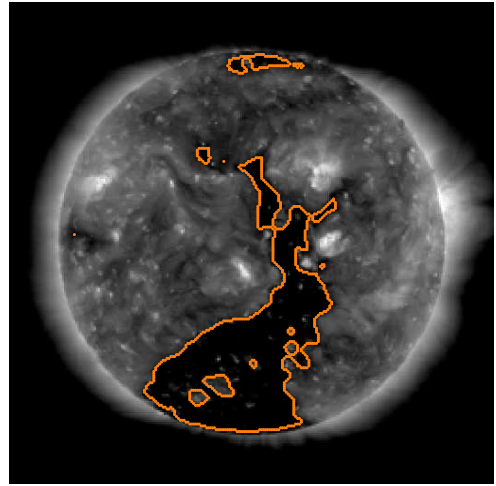}
    \caption{Contours of coronal holes isolated by U-Net (January~30,~2017). Here and below image size is $256\times 256$ pixels.}
    \label{fig:ch_map}
\end{figure}

The trained network was used to predict CHs in the test part of the dataset. Recall that the test part 
consists of daily SDO/AIA~193~{\AA} images from January 2017 till June 2018. The Fig.~\ref{fig:ch_map} 
shows an example of CHs segmentation (that day, January 30, 2017, U-Net found the largest CH in the
test period, about 13\% of the solar hemisphere).

We do not apply any post-processing for obtained segmentation maps
due to the following reasons. First, we want to demonstrate a
baseline for CHs segmentation with neural network, i.e. a quality that
can be expected from the neural network trained just on original images
and target segmentation maps.
Second, we believe that most of drawbacks in predicted outputs
can be improved not due to hand-engineered corrections, but due to improvements in the network
architecture or in the network training scheme. Third, 
in our opinion, comparison of CHs against magnetograms and 
discarding of some CHs with respect to their unipolarity is
a very speculative procedure and requires a detailed
investigation, which is out of the scope of this paper.

We appreciate that CHs isolated with the above mentioned procedure
are rather only candidates to CHs. However, since
the ground truth is unknown, we will concentrate 
on demonstration of stability of 
the procedure and its comparison with alternative approaches.
For comparison we consider catalogue of CH isolated by CHIMERA (available at \href{https://solarmonitor.org/data/}{https://solarmonitor.org/data/}, by SPOCA (available at \href{http://lmsal.com/isolsearch}{http://lmsal.com/isolsearch} and via the Application Programming Interface (API) of the Heliophysics Events Knowledgebase (HEK)) and
CH maps from test part of our dataset. We failed to find any other
public available long-term catalogue of isolated and digitized CHs boundaries 
and its parameters.
It would be fruitful to include into investigation CH maps from
the National Oceanic and Atmospheric Administration (NOAA) Space Weather Prediction Center (SWPC) and
CH maps from Automatic Solar Synoptic Analyzer (ASSA). However, the first ones are not digitalized
(i.e. CHs boundaries are not extracted), the second ones include only latest map.

\subsection{Dice score as a similarity measure}
We will compare CH segmentation maps using the Dice similarity
coefficient \citep{dice}. For two sets $A$ and $B$ it is expressed as
\begin{equation}
dice(A, B) = \frac{2 |A \cap B|}{|A| + |B|} \, ,
\end{equation}
where $|A|$ is the cardinal of set $A$. Intuitively, 
this can be seen as a percentage of overlap between the two sets.

\begin{figure}
	\includegraphics[width=\columnwidth]{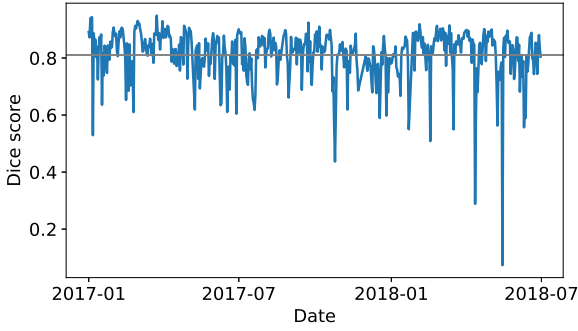}
    \caption{Dice scores between segmentation maps predicted with U-net and obtained with the region
growth algorithm for the test time period.
Horizontal line shows mean value 0.81.}
    \label{fig:dice}
\end{figure}

First, we compare CH segmentation maps predicted with U-net with CH maps for the same day obtained with the region growth algorithm. 
Fig.~\ref{fig:dice} shows the dice score between
two segmentation maps for the test time period. 

We found the dice score fluctuates moderately near its mean values which is equal to 0.81 with standard
deviation equal to 0.1. To better
understand this value it can be noted that
two concentric circles with outer radius 20\% greater than inner have the dice score 0.82. 

\begin{figure}
\includegraphics[width=\columnwidth]{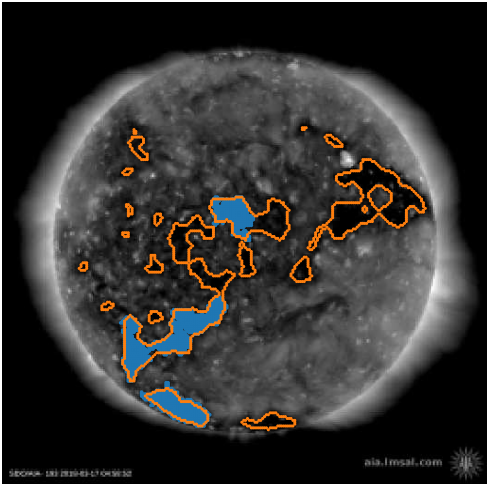}
    \caption{Substantial difference between
    CHs isolated by the region
    growth algorithm (blue regions) and by U-net (orange contours). March 17, 2017.
    Dice score is 0.55.}
    \label{fig:diff_large}
\end{figure}

Visual comparison of predicted and target segmentation maps brings us to conclusion that U-Net finds more regions related to CHs and gives more accurate approximation
of them. Rare outliers in Fig~\ref{fig:dice} are mostly caused
by substantial underestimating of CH regions by
the region growth algorithm. An example of the outlier with
dice 0.55 is shown in Fig.~\ref{fig:diff_large}. Again, we observe that U-Net provides better segmentation result.

\begin{figure}
\includegraphics[width=\columnwidth]{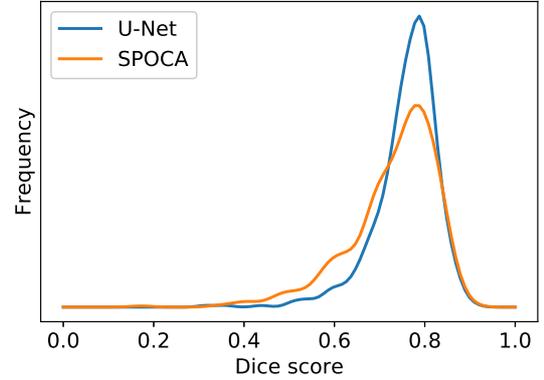}
    \caption{Distribution of dice scores
    for segmentation maps produced by U-Net (blue line) and SPOCA (orange line).}
    \label{fig:dice_distr}
\end{figure}

Dice score also can be used to validate a stability of
segmentation maps. 
Indeed, considering CH segmentation maps
not in the solar disk, but in the Carrington 
coordinate system, we can naturally expect that images 
do not vary a lot from day to day. Thus, it makes sense
to compute the dice score. To neglect the rotation effects
we consider the dice score between current map and union of maps for the day before and after the current day.
Then we compute a histogram of dice scores obtained for the test time period.
Fig.~\ref{fig:dice_distr} shows the smoothed histograms for U-Net and SPOCA. 

One can note in the Fig.~\ref{fig:dice_distr} that while modes of the distributions are very close, frequency for U-Net is about 25\% higher.
This means, that U-Net produces more consistent segmentation maps and demonstrates better stability as compared to SPOCA.
Unfortunately, catalogue of CH isolated by CHIMERA does not contain boundaries of CH and thus is not
included into this analysis.

\subsection{Time-variation of the CHs areas}

Here we consider another aspect of stability of CHs
segmentation algorithm, which is a time-variation of
total CHs area. Since typical CHs exist more than one solar rotation period, it is highly unlikely to observe rapid
day-to-day leaps in a graph of CHs area. 

Recall that one can calculate areas of solar active regions in sky plane and measure them in percentage of solar disk or, alternatively, estimate reprojected areas and measure them in percentage of solar hemisphere. In practice, however, estimation of reprojected areas can be inaccurate since
many CHs are located near poles where projection effects are substantial. On the other hand, ares measured in sky plane
can also be misleading since two CHs of the same area (in sky plane) located near disk centre and near pole have very different actual (reprojected) areas. Below we will show areas calculated in both ways.

\begin{figure*}
\includegraphics[width=\textwidth]{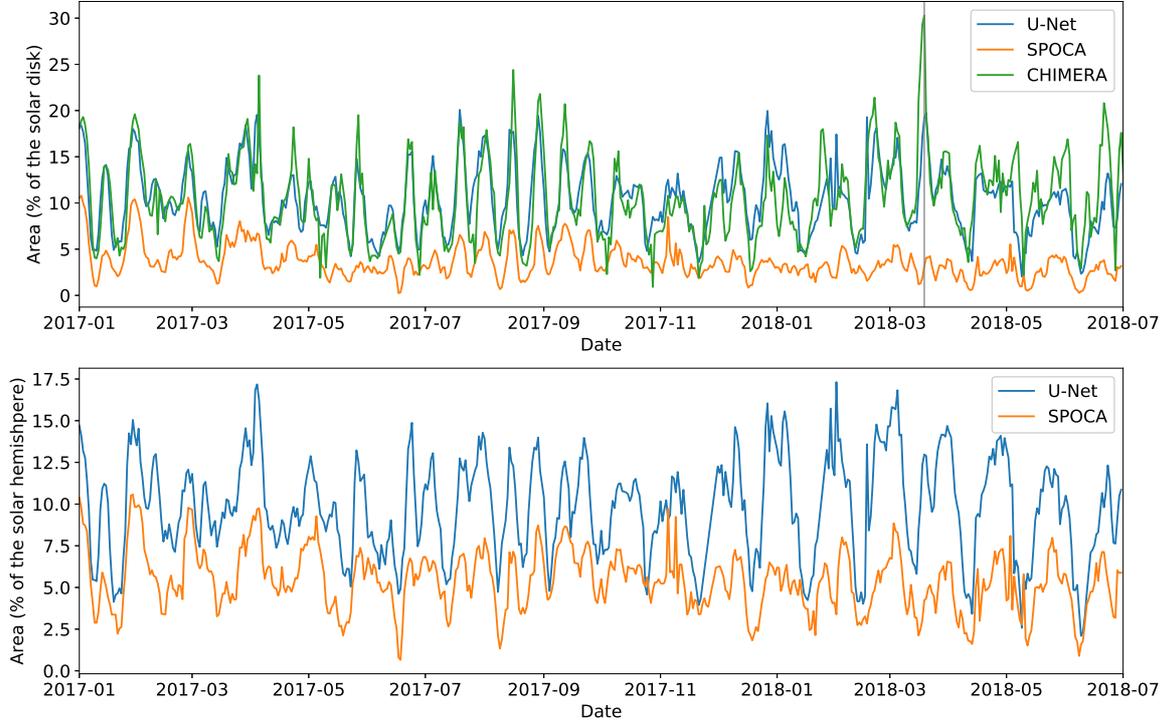}
    \caption{Upper panel: total area of coronal holes in plane of sky. Blue line is for U-Net, orange line if for SPOCA,
    green line is for CHIMERA algorithm of CHs segmentation.
    Vertical line shows a day selected for detailed comparison (see Fig.~\ref{fig:19march}).
    Lower panel: total reprojected area of coronal holes.
    Blue line is for U-Net, orange line if for SPOCA
    algorithm of CHs segmentation.}
    \label{fig:area}
\end{figure*}

\begin{figure*}
\centering\includegraphics[width=\textwidth]{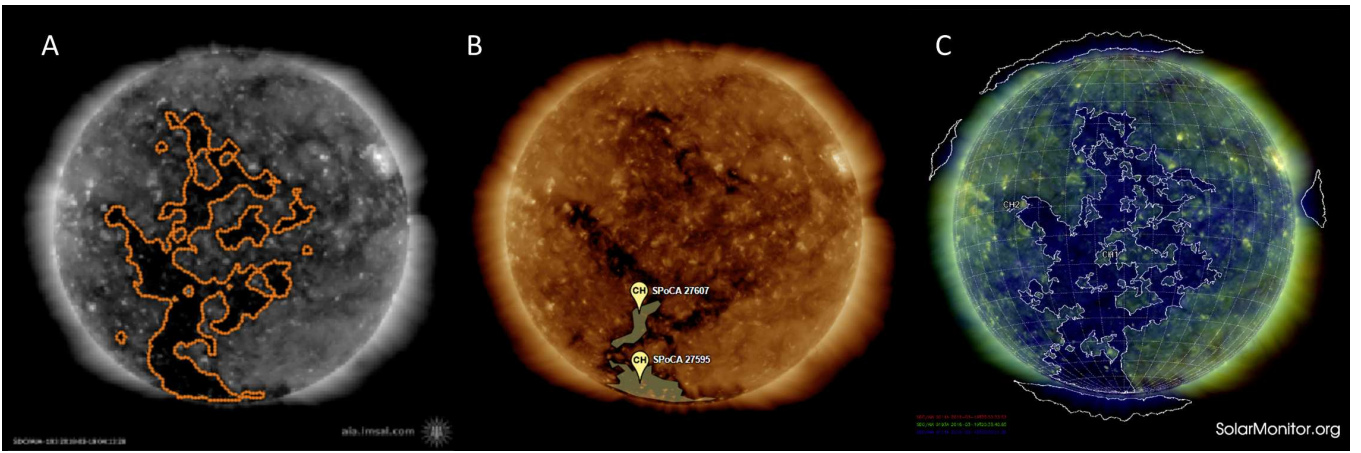}
    \caption{Segmentation maps for March 19, 2018 by U-Net (A),
    SPOCA (B) and CHIMERA (C).}
    \label{fig:19march}
\end{figure*}

Fig.~\ref{fig:area} (upper panel) shows a daily variation of total CHs area
measured in sky plane. We observe that U-Net and CHIMERA
give very similar results with a correlation coefficient equal to 0.76. 
At the same time, total CHs areas according
to SPOCA are up to 3 times fewer in contrast
to CHIMERA and U-Net. We selected a day (March 19, 2018) when the difference is
most prominent and show isolated CHs corresponding to each algorithm in Fig.~\ref{fig:19march}. In our opinion, the map B (SPOCA) clearly underestimates
CHs regions, the map C (CHIMERA) somehow 
overestimates CHs with respect to the 193~{\AA} image, while the map A (U-Net) produces
the most accurate segmentation that corresponds to visual
expectation. Note that total CHs area shown in Fig.~\ref{fig:19march} is 19\% of the
solar disk according to U-Net, 3\% for SPOCA and 30\% for CHIMERA. 

In Fig.~\ref{fig:area} (lower panel) we show a daily variation of total reprojected CHs area.
It does not include corresponding plot by CHIMERA since we observe that almost all reprojected areas in catalogue of CHs isolated by CHIMERA are in order of magnitude larger than that for SCOCA or U-Net. The point is that according to the source code at \href{https://github.com/TCDSolar/CHIMERA}{https://github.com/TCDSolar/CHIMERA}
CHIMERA estimates reprojected area relative to position of CHs centroid. In our opinion,
this method can make sense for small objects, i.e. pores, while for large object, i.e. CH, the more accurate estimation requires e.g. reprojection of each pixel within the CH. For example,
consider a day March 9, 2018. According to CHIMERA, the largest CH near the north pole occupies 2.7\% of the solar disk area and has reprojected area $1.3 \cdot 10^6$ ${\rm Mm}^2$.
However, assuming solar radius equal to $6.95 \cdot 10^2$ Mm we obtain that this CH should cover $1.3 \cdot 10^6 / 2\pi (6.96 \cdot 10^2) ^2 \cdot 100\% = 42.7\%$ of the solar hemisphere, which does not correlate with visual expectation and is a clear overestimation. For comparison, corresponding CH isolated by U-Net has an area 3.5\% of the solar disk and 4.6\% of the solar hemisphere.

Considering Fig.~\ref{fig:area} we observe that areas by SPOCA are 
about half of areas by U-Net, but both demonstrate similar 
variations with correlation coefficient 0.66.

\subsection{Difficult cases}

\begin{figure}
\includegraphics[width=\columnwidth]{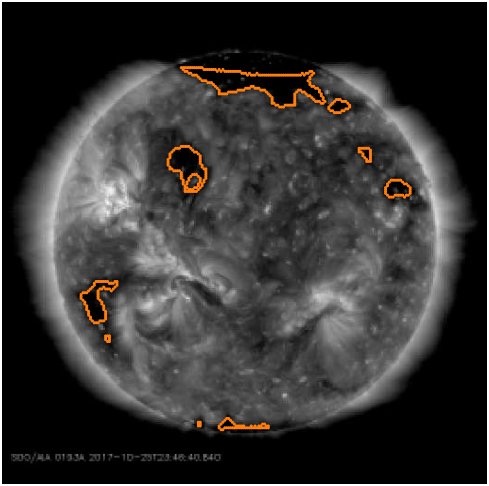}
    \caption{Almost absent solar limb can result in incorrect CH segmentation (see the CH near the north pole).}
    \label{fig:ch_no_mask}
\end{figure}

We have noted several occasional cases, when U-Net predicts unnatural CHs. 
They occurs when the solar limb separating
CH and the outer space is almost absent, e.g. as one near the north pole shown in the Fig.~\ref{fig:ch_no_mask} for October 25, 2017. 

\begin{figure}
\includegraphics[width=\columnwidth]{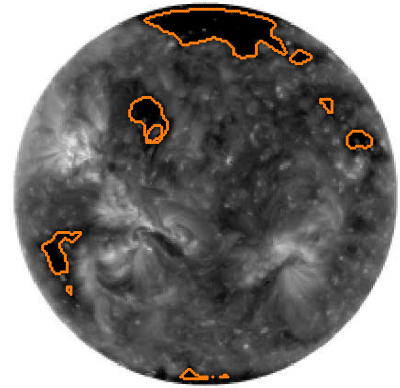}
    \caption{Prior setting the maximal intensity to pixels
    outside the solar disk improves the CH segmentation near the solar limb in comparison to the Fig.~\ref{fig:ch_no_mask}}
    \label{fig:ch_with_mask}
\end{figure}

The point is that the extremely thin solar limb can not propagate 
deeper into U-Net layers
and fire neurons responsible for CH detection. 
However, 
one can easily improve the situation just by a prior segmentation of the solar disk. In the Fig.~\ref{fig:ch_with_mask} we set 
maximal intensity to pixels outside the solar disk.
As we can see, now U-Net isolates CH near the north pole correctly.

Thus, to avoid the limb effects we recommend including 
a segmentation of the solar disk in image preprocessing pipeline.

\subsection{CHs in the solar cycle 24}
Here we present a variation of CH areas
from the beginning of SDO/AIA observations in 2010
up to the present time according to U-Net
segmentation. Note that this time period includes both
train and test periods for U-Net and covers approximately the solar cycle 24. For comparison
we consider a variation of the solar
wind (SW) speed measured onboard the Advanced
Composition Explorer \citep[ACE][]{Stone1998}. Daily
averaged SW speeds were obtained from level 2
ACE data at \href{http://www.srl.caltech.edu/ACE/ASC/level2/}{http://www.srl.caltech.edu/ACE/ASC/level2/}.

\begin{figure*}
\includegraphics[width=\textwidth]{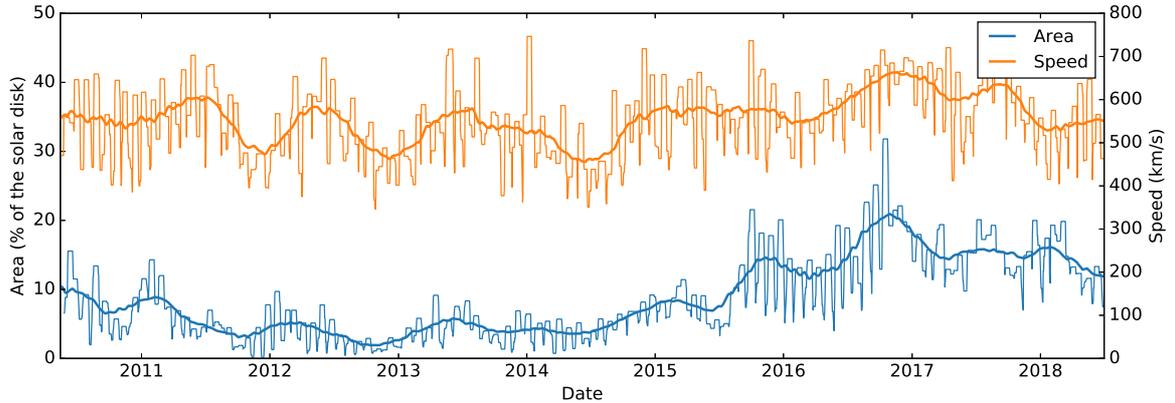}
    \caption{Maximal within 13-day period areas of coronal holes (blue line) and solar wind speed (orange line). Smoothed lines within each plot represent 150-day moving average.}
    \label{fig:ace}
\end{figure*}

Due to the fact that visible CH areas vary with
the solar rotation, some data aggregation is required
to reduce this effect. We replace total CH area for the current day 
with the maximal CH area within the 13-day window centred at the current day. 
Note that 13-day window roughly corresponds to a half of the solar rotation period. 
The same procedure is applied for SW speed data as well.
The result is shown in the Fig.~\ref{fig:ace}.
Additional 150-day moving average, also shown in the Fig.~\ref{fig:ace}, reveals
long-term variations of the obtained time-series.

We observe that CHs area demonstrates yearly variations,
increasing trend during the declining phase of the solar cycle
and is minimal during the maximum of the solar cycle.
In our opinion, the nature of near-yearly variations can be in North-South asymmetry and variations of the solar D-angle,
while increasing of CHs area in the minimum of the solar cycle is associated with polar magnetic field strengthening. Note that CH areas vary in about
three times within the solar cycle. We also note a correlation between averaged CH areas and averaged SW speeds.
Numerical value of this correlation is 0.7. More detailed investigation of the coupling between CHs and SW can be found e.g. in \citet{Rotter2015} and \citet{deToma2011}.

\section{Conclusions}
Segmentation of coronal holes is a basic step in many
space weather prediction models. However, as we have demonstrated,
isolated CHs and their parameters vary dramatically from
one algorithm to another one. In our opinion, the problem
comes from the fact that hand-engineered algorithms do
not have enough capacity to deal with a large variability of CHs. 
To overcome this limitation, we suggest
an approach based on modern neural network architecture, known as U-Net,
which has approved its effectiveness in various image
segmentation problems.

We trained the U-Net neural network on a set of daily SDO/AIA~193~{\AA}
solar disk images and corresponding CH segmentation maps for the time period from 2010 to 2017. Segmentation maps were
provided by the Kislovodsk Mountain Astronomical Station. 
The time period from 2017 to June 2018 was used for model evaluation 
and comparison with other segmentation algorithms.

The source code that allows to reproduce the model architecture and training
procedure is available at \href{https://github.com/observethesun/coronal_holes}{https://github.com/observethesun/coronal\_holes}.

Our first observation is that the trained neural network
gives a better prediction of CHs in contrast to
the semi-automatic procedure applied for CH segmentation
in the training dataset. Thus, we conclude that the neural
network is able to generalize and improve heuristics that
were exploited for annotating of the training dataset.

Second, detailed comparison of CHs isolated by U-Net, SPOCA and CHIMERA 
algorithms shows that CHs isolated by U-Net correlate better with
visual expectation from SDO/AIA~193~{\AA} solar disk images. 
U-Net also provides better stability of segmentation 
maps in comparison to SPOCA.

Third, we observe very similar patterns in daily variation
of total CHs area between all three algorithms, however,
absolute values differ a lot.

We conclude that U-Net is able to give
a reasonable segmentation of CHs in solar disk images.
The most important advantage of this approach is that
neural networks can learn and generalize a methodology
of active region isolation even if this methodology is
not formalized (e.g. in case of manual segmentation).
Thus one can create extended homogeneous datasets of active regions.

As an example of homogeneous dataset we considered
CHs isolated by U-Net from the beginning of SDO/AIA observations in 2010
(solar cycle 24). We observed that CHs area increases in about three times
from minimal values in the maximum of the solar cycle to maximal values
during the declining phase of the solar cycle. Comparison with the solar wind speed
variations for the same period gives the correlation coefficient 0.7.

We hope that this work will inspire further investigations
of neural networks in application to analysis of solar images.

\section*{Acknowledgements}
We are grateful to the referee for critical reading of the manuscripts.
The research is supported by RSF under grant 15-12-20001 (AT)
and RFBR under grant 18-02-00085 (EI).

\bibliographystyle{mnras}

\bsp	% typesetting comment
\label{lastpage}
\end{document}